\documentclass[journal]{vgtc}                     

\newcommand{\new}[1]{#1} 

\onlineid{1666}

\vgtccategory{Research}

\vgtcpapertype{theory/model}

\title{Deconstructing Implicit Beliefs in Visual Data Journalism: Unstable Meanings Behind \textit{Data as Truth} \& \textit{Design for Insight}}

\author{%
  \authororcid{Ke Er Amy Zhang}{0009-0006-9132-006X},
  \authororcid{Jodie Jenkinson}{0000-0002-4066-9732}, and 
  \authororcid{Laura Garrison}{0000-0001-7134-2006}
}

\authorfooter{
  \item
  	Ke Er Amy Zhang and Laura Garrison are with Univ. of Bergen.
  	E-mail: \{ke.zhang\,$|$\,laura.garrison\}@uib.no\,.
  \item
  	Jodie Jenkinson is with Univ. of Toronto.
  	E-mail: j.jenkinson@utoronto.ca\,.

}

\abstract{%
 We conduct a deconstructive reading of a qualitative interview study with 17 visual data journalists from newsrooms across the globe. We borrow a deconstruction approach from literary critique to explore the instability of meaning in language and reveal implicit beliefs in words and ideas. Through our analysis we surface two sets of opposing implicit beliefs in visual data journalism: \textit{objectivity/subjectivity} and \textit{humanism/mechanism}. We contextualize these beliefs through a genealogical analysis, which brings deconstruction theory into practice by providing a historic backdrop for these opposing perspectives. Our analysis shows that these beliefs held within visual data journalism are not self-enclosed but rather a product of external societal forces and paradigm shifts over time. 
 Through this work, we demonstrate how thinking with critical theories such as deconstruction and genealogy can \new{reframe ``success'' in visual data storytelling} and diversify visualization research outcomes. These efforts push the ways in which we as researchers produce domain knowledge to examine the sociotechnical issues of today's values towards datafication and data visualization. All supplemental materials for this work are available at \href{https://osf.io/5fr48/}{\texttt{osf.io/5fr48}}.
}


\teaser{
  \centering
  \includegraphics[width=\linewidth, alt={An isometric cube, lower left, decomposes into subcubes that form trails threading across the branches of a genealogy tree overlaid with historical and contemporary visualizations.}]{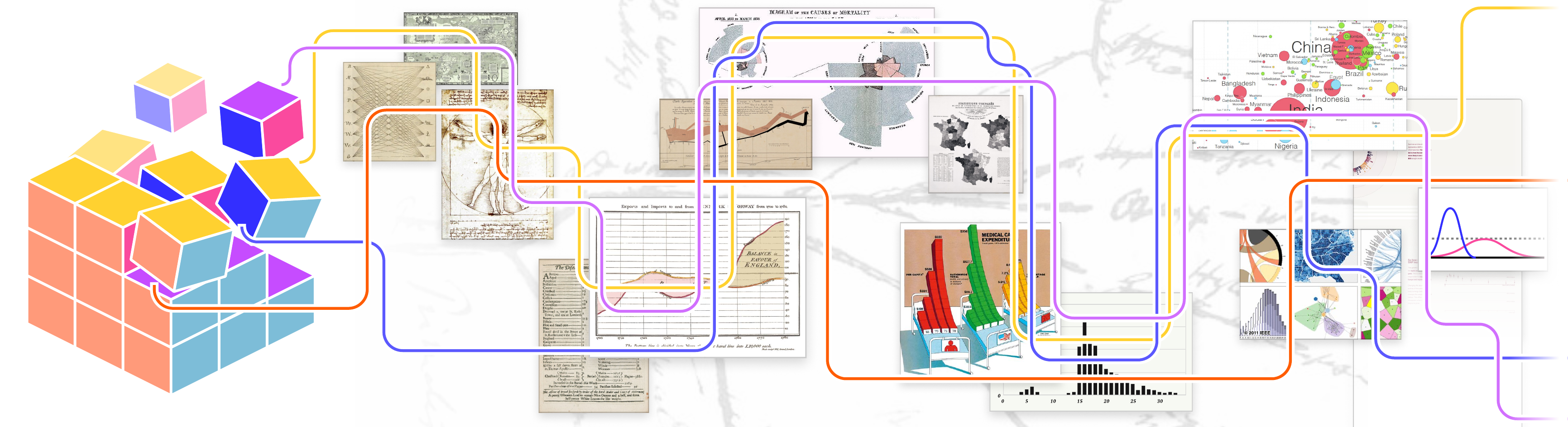}
  \caption{
    A conceptual overview of our analysis that (1) \textit{deconstructs} implicit beliefs and tensions amongst visual data journalists that we interviewed and (2) follows the \textit{genealogy} of these tensions from the Renaissance to present day. 
  }
  \label{fig:teaser}
}

\graphicspath{{figs/}{figures/}{pictures/}{images/}{./}} 

\usepackage{tabu}                      
\usepackage{booktabs}                  
\usepackage{lipsum}                    
\usepackage{mwe}                       
\usepackage{soul}
\usepackage{nowidow}
\usepackage{pifont}
\usepackage[svgnames]{xcolor}

\usepackage{mathptmx}                  

\renewcommand{\colorbox}[2]{\colorlet{tmpcolor}{#1}\sethlcolor{tmpcolor}\hl{#2}}

\renewenvironment{quote}{%
  \list{}{%
    \leftmargin0.3cm   
    \rightmargin\leftmargin
  }
  \item\relax
}
{\endlist}

\definecolor{objCol}{HTML}{CBCDFF}
\definecolor{subjCol}{HTML}{E5E9FF}
\definecolor{humCol}{HTML}{FFB9DB}
\definecolor{mechCol}{HTML}{FFE0EF}

\newcommand{\obj}[1]{{\setlength{\fboxsep}{0pt}\colorbox{objCol}{#1}}} 
\newcommand{\subj}[1]{{\setlength{\fboxsep}{0pt}\colorbox{subjCol}{#1}}} 
\newcommand{\hum}[1]{{\setlength{\fboxsep}{0pt}\colorbox{humCol}{#1}}} 
\newcommand{\mech}[1]{{\setlength{\fboxsep}{0pt}\colorbox{mechCol}{#1}}} 

\DeclareUnicodeCharacter{2729}{\ding{72}}

\begin{document}


\firstsection{Introduction}
\label{sec:introduction}
\maketitle
Visual data journalism is an established medium to support readers in navigating the news through statistics and stories about data~\cite{antonopoulos2020data}. 
Data journalists' workflows closely mirror the narrative visualization pipeline~\cite{lee2015more}: data gathering, cleaning, filtering, then strategically mapping visual variables in a sequential narrative that tells a compelling story~\cite{segel2010narrative, weber2012data, showkat2021stories, bradshaw2011}. 
Much work in visualization has focused on formalization~\cite{segel2010narrative, hullman2011visualization, lee2015more}, tools to facilitate data-driven storytelling~\cite{shi2020calliope, morini2023shock}, or the effects of design and interaction choices on readers~\cite{garreton2023attitudinal, kauer2025towards}. 

While many works focus on data journalists as domain experts or target users, comparatively less research probes into the \textit{values} and \textit{beliefs} that drive journalists' design and interaction choices, or the underlying norms and beliefs around data storytelling. Research investigating the way that data journalists think~\cite{weber2012data, showkat2021stories} typically focuses on constructivist, e.g., grounded theory~\cite{glaser1967grounded} or interpretivist paradigms, e.g., reflexive thematic analysis~\cite{braunthematic2021}. 
While these qualitative methods enable the researcher to bring their own experience and context to the analysis to form new understandings of practice, their design only affords analysis of that which is \textit{explicit}, e.g., what is said, what is shown~\cite{jackson2011thinking}. These analysis methods furthermore are predicated on assumptions of structures of \textit{stable} interrelations that shape human perception and experience, e.g., the stability of semiotics in visual--data mappings~\cite{bertin1983semiology}, or narrative sequence distilled to \textit{the Hero's Journey}~\cite{campbell2008hero}.    
\new{Critical theory challenges} these normative assumptions and considers subtext as part of the analysis\cite{tyson2023critical}. Applied to visualization, feminist~\cite{barad2018diffracting, akbaba2023troubling, akbaba2024entanglements}, humanist~\cite{drucker2014graphesis, drucker2017information}, and genealogical~\cite{foucault2003essential, correll2024body} inquiry allows the researcher to question the sociocultural, technical, and historical structures that impact values in visualization design. 
Also under the umbrella of critical theory is \textit{deconstruction}, an approach \new{often applied in literary criticism} that assumes all language is unstable. 
This perspective allows the researcher to identify implicit beliefs and value tensions in a text~\cite{derrida1974grammatology, rivkin2017literary, tyson2023critical}. While \new{constructivist or interpretivist} analysis asks, \textit{What do journalists believe are core elements of engaging narratives?}, a deconstructive analysis asks, \textit{What norms or standards impact data journalists' beliefs about \new{success} and meaning in visual data journalism?} 

We see a critical need to ask such questions, particularly for visual data journalism. In our \textit{Golden Age of Data Visualization}~\cite{marriott2024golden} we face socioethical concerns around datafication that include threats to privacy, exploitation and discrimination of socially vulnerable populations, rationalization and rhetorical power of data, and the affective result of living with a constant stream of information. Visual data journalism is uniquely positioned at the tension of these issues; a field trained in and aware of the use of narratives and rhetoric, yet operating with a code of ethics pursuing truth, transparency, and rigour~\cite{showkat2021stories}.

To examine these tensions and their relations to norms or standards driving the data-driven news stories produced today, we conducted interviews with 17 visual data journalists that included consultants, newsroom staff, and editorial directors. Through our analysis \new{(conceptually summarized in \cref{fig:teaser})} of these interviews, we report on journalists' explicit and implicit notions of \textit{data as truth} and \textit{design for insight} in visual data journalism. Through deconstructive analysis of interview transcripts we dismantle these views and the \textit{objectivity/subjectivity} and \textit{humanism/mechanism} dialectics that they rest upon. Through a genealogical analysis we underline the difficulty journalists face in challenging \new{these dialectics}: data as \textit{objective} truth has been entrenched in Western canon for centuries; 
design for \textit{humanistic} insight is simultaneously more recent and older than \textit{mechanistic} design practice. 
We observe in our participants a deep commitment to quality and empathetic reporting that rests on a \new{set of beliefs} that collapses historically separate binary \new{oppositions} of \textit{data as objective/subjective truth} and \textit{design for humanistic/mechanistic insight}. The beliefs are entrenched in broad sociocultural, technical, biopolitical contexts that support existing power structures~\cite{foucault2003essential}. Freedom from these entrenched beliefs affords \textit{explicit} reflection on the ways that \textit{implicit} beliefs in data journalism, and visualization at large, drive design choices that expose vulnerable groups through well-intended efforts towards transparency and engagement. We can then better decide which stories are right to tell, and how to make visible these complex, dynamic contexts. 

\textbf{Our primary contribution is the surfacing of implicit, unstable epistemic beliefs held in visual data journalism.} 
Through a deconstructive analysis~\cite{derrida1974grammatology, tyson2023critical} of our interview transcripts we expose a set of \textbf{binary oppositions}, i.e., \new{contradictory philosophical beliefs,} in visual data journalism: \textit{objectivity/subjectivity} and \textit{humanism/mechanism}. We follow with a genealogical analysis~\cite{foucault2003essential, akbaba2024entanglements, correll2024body} that contextualizes the privileging of one belief over the other to show that these beliefs are not self-enclosed but rather a product of external, complex sociocultural and political forces through history. \new{We discuss the implications of these beliefs on visual data journalism as well as visualization at large.}
\textbf{Our secondary contribution is an introduction to deconstruction as a method for qualitative interview analysis in visualization research.} We introduce its theoretical underpinnings and show its application from literary analysis to reframe our research questions to examine implicit beliefs as products of sociocultural events and paradigm shifts, rather than as fixed structures. 

\section{\new{Terminology}}
\label{sec:terms}

Our work uses several terms and concepts from comparative literature and philosophy that are either new or ambiguously used in visualization. Inspired by Akbaba et al.~\cite{akbaba2024entanglements}, this section introduces and clarifies our use of these terms, particularly \textit{poststructuralism}, \textit{deconstruction}, and \textit{genealogy}. Additionally, we contrast \textit{objectivity} with \textit{subjectivity}, and \textit{mechanism} from \textit{humanism}. 

\textit{Poststructuralism} is a movement rooted in 1960s France which asserts that our social, cultural, and political systems of meaning are inherently ambiguous~\cite{harcourt2007answer,howarth2013poststructuralism}. Poststructuralism is closely related to and overlaps significantly with \textit{critical theory}, which has a similar project~\cite{dork2013critical, tyson2023critical}. Critical theories challenge the norms and assumptions of technology that we often take as given on society, e.g., race~\cite{ogbonnaya2020critical}, gender~\cite{bardzell2010feminist}, and class~\cite{sharma2023post}. These theories offer a method of interrogating the impact of historical and sociocultural events on today's technological practices~\cite{bodker2023re}. While \textit{structuralism} assumes `universal' laws and behaviours through stable systems of relationships~\cite{hawkes1977structuralism}, \new{e.g., visualization design spaces~\cite{bach2018narrative} and classifications~\cite{pokojna2024language}}, poststructuralism rejects such assumptions to probe the nature of human subjectivity and how context shapes identity and meaning in relation to structure, power, and agency~\cite{howarth2013poststructuralism}, \new{e.g., power relations in visualization through enunciation theory~\cite{drucker2017information}}. \textit{Deconstruction} is a theory within poststructuralism that is best known in literary critique~\cite{derrida1974grammatology, derrida2004dissemination, rivkin2017literary} and later in the visual arts~\cite{Brunette1994}. We use deconstruction in the literary sense to expose the contradictory ideologies, norms, and assumptions that are implicit within a text~\cite{jameson2024years}. Connected to deconstruction through poststructuralism is \textit{genealogy}~\cite{harcourt2007answer}, which we use in reference to the analysis strategy described by French philosopher Michel Foucault in the 1970s~\cite{foucault2003essential}. Genealogy threads a concept or idea through historical events, people, and places to contextualize it among the networks of external societal forces and institutional power dynamics~\cite{tamboukou1999writing, foucault2003essential}.
We apply deconstructive and genealogical analysis to interview transcripts to understand the dominant \textit{epistemologies}, i.e., standards and practices that drive knowledge production, \new{e.g., alternate visualization epistemologies~\cite{akbaba2024entanglements}}. 

Later in Sec.~\ref{sec:deconstruction} and \ref{sec:genealogy} we discuss two binary perspectives: \textit{objectivity/subjectivity} and \textit{humanism/mechanism}. Each of the terms within these oppositions are epistemic virtues, i.e., behaviours or personal qualities that support acquiring and applying knowledge~\cite{daston2021objectivity}. We characterize \textit{objectivity} within the frame of modern science: an external observation of events where information is gathered and predictions are made through experimentation and mathematical formalization~\cite{gillispie2016edge}, \new{e.g., algebraic model of visualization~\cite{kindlmann2014algebraic}}. \textit{Subjectivity}, then, considers events through the lens of personal experience and perspective~\cite{gillispie2016edge, daston2021objectivity}, \new{e.g., diffractive reading~\cite{akbaba2023troubling}}.  
In our references to \textit{mechanism} and \textit{mechanistic}, we mean deterministic, algorithmic, or mechanical qualities that enhance productivity~\cite{daston2021objectivity}, \new{e.g., algorithmically-generated data stories~\cite{shi2020calliope}}. In contrast, we use \textit{humanism} and \textit{humanistic} in reference to the philosophical position that centres human activities, thought, and emotion, \new{e.g., affective visualization design~\cite{lan2023affective}}. 

\section{Related Work}
\label{sec:relatedwork}

In this section, we summarize related work on visual data journalism and critical perspectives in visualization research that inspire our work. 

\subsection{Practices in visual data journalism}

Data-driven storytelling in the newsroom and data visualization agencies innovate on the visual form of a story to communicate a variety of topics to diverse stakeholder groups. Dynamic and interactive data visualizations have become staples supporting news stories about public health~\cite{times2021coronavirus}, economics~\cite{economist2017globalization}, and politics~\cite{nytimes2024needle}. Design innovations from within the newsroom have dramatically changed the landscape of visual narrative in the public sphere, such as pushing the boundaries of article-based multimedia through the scrollytelling genre~\cite{nytimes2012snowfall} or the publication of data visualization libraries like D3~\cite{d2023data}. 
Visualization research is equally fascinated with data-driven storytelling, often drawing from and analyzing newsroom graphics to conceptualize data-driven narratives by developing frameworks~\cite{segel2010narrative, hullman2011visualization, lee2015more}, processes~\cite{lee2015more}, guidelines~\cite{meuschke2022narrative}, and design spaces~\cite{bach2018narrative, hao2024design}. 
Work in this tradition usually takes a primarily structuralist perspective in furthering our understanding of narrative design. Specific work to support data journalists' workflows~\cite{shi2020calliope, sun2022erato} distills the complex and interwoven space of visual story design to its essential parts for efficiency and insight. Other approaches to support data journalists investigate critical visualization strategies such as plurality~\cite{kauer2025towards}, empowerment~\cite{garreton2023attitudinal}, or contingency~\cite{morini2023shock} that hint towards alternative ways of seeing and knowing. Evaluating reader response to novel research artifacts is equally important to the development of the tools themselves, with increasing research investigating affect and engagement with new approaches~\cite{yang2023swaying, lan2023affective}.
Further work by Yang et al.~\cite{yang2024backstory} has brought a new dimension to this space by reflecting on the visualization designer's choices in creating election visualizations that, similar to our work, seeks opportunities for conversation about visualization for complex topics in the public sphere. 
With data-driven storytelling on the rise in society~\cite{engebretsen2020data} and various techniques and libraries available to facilitate data journalistic endeavours, increasing visualization works investigate the ways that data journalists work and think. Research along this line includes, for example, specific approaches to visualizing electoral processes~\cite{cai2024watching}, strategies for creating and identifying newsworthy narratives~\cite{weber2018data, engebretsen2020data, showkat2021stories}, and broader analyses on the evolving role of data visualization in journalism alongside the other tasks of a data journalist~\cite{engebretsen2017visualization, fu2023more}. Perhaps most similar to our work, Dhawka \& Dasgupta~\cite{dhawka2025social} critically examine journalist beliefs and bias in the design process, although they employ an interpretive analysis process to analyze issues of race and gender. In contrast, we use a different analysis process to uncover hidden ideological contradictions.

\subsection{Critical approaches to visualization}

Critical visualization, first conceptualized as a set of principles that examine issues of author--reader agency hidden within visualizations~\cite{dork2013critical}, is now broadly understood as the use of critical theory to examine power structures in visualization practice. Through a humanistic lens, Drucker unpacks the norms and standards that have shaped visual epistemology at large~\cite{drucker2014graphesis}, advocating for researchers to take ``critical insights from literary, cultural, and gender studies'' to invigorate visualization design. 
To date, feminist epistemology has been primarily drawn upon to understand visualization design practices from a new lens, i.e., differing from traditional behavioural and cognitive methods~\cite{cleveland_graphical_1984}. D'Ignazio \& Klein~\cite{d2023data} apply feminist theory to data visualization to produce a set of pragmatic principles that help researchers account for non-neutrality, power and privilege in their practice. Akbaba et al.~\cite{akbaba2023troubling} continues to show how theory can be used productively as methodology, demonstrating diffraction~\cite{lazar2021adopting, barad2018diffracting} as an interview analysis strategy and offering alternative visualization epistemology~\cite{akbaba2024entanglements} borrowing from feminist entanglement theory~\cite{barad2007meeting}. Outside of feminist theory, issues of power in visualization culture have been examined through poststructural approaches, e.g., power relations through enunciation theory~\cite{drucker2017information} and historical data cultures through genealogy~\cite{correll2024body}. In our search for methods to support interrogation of belief structures, we found that deconstruction, which questions the inherent stability of structures in language and visuals, has been sparsely used in visualization and greater HCI. Bertschi~\cite{bertschi2007without} proposes a theoretical foundation for knowledge visualization built on deconstruction, while Chiasson \& Davidson~\cite{chiasson2012reconsidering} demonstrate a deconstructive reading of texts within information systems. We did not find other demonstrations of deconstruction beyond the examples above and see this as motivation to investigate what deconstruction can bring to qualitative analysis in visualization research.

\section{Interview Methods}
\label{sec:methods}

Our initial idea for this study followed a structuralist approach: to derive a visualization design space for engaging narratives that dealt with socially critical topics. However, as we began planning and preparing for this study, our aims rapidly transformed into poststructural questions on the \textbf{standards of practice and beliefs that lead data journalists to work within their existing design space.} We felt that identifying the data journalist's design space told only a small part of the story; instead, we thought that starting with learning how data journalists \textit{think} about how to engage readers would ultimately result in a richer, more nuanced understanding for \textit{why} and \textit{how} public-facing visual news stories are made. In this section, we describe our protocol to recruit and interview participants. We provide some details on participants as well as our transcription and anonymization process for the interview data.

\subsection{Participants}
 We used purposive sampling~\cite{taherdoost_sampling_2016} to recruit visual data journalists with experience crafting visual data stories for newsrooms. We used this method to support recruitment of a diverse set of data journalists working on a range of topics in different work settings, geographic, and cultural contexts \new{to achieve a rich, in-depth qualitative interview study~\cite{hennink2022sample}.} To ensure that we recruited individuals working in diverse styles, topics, and projects, we first identified exemplars of digital news stories and articles that displayed principles of narrative visualization, e.g., use of different narrative genres, degree of author--reader agency~\cite{segel2010narrative}. In identifying data stories spanning a broad range of style and perspective, we also sought to include work embodying critical approaches in visualization design~\cite{dork2013critical}: \textit{disclosure} (show decisions made about data, representation, and interaction), \textit{plurality} (show more than one facet of a phenomenon to support different interpretations), \textit{contingency} (allow for different ways to experience a visualization), and \textit{empowerment} (allow for viewers to question visual representations and use these representations to tell their own story, e.g., through interaction). We expanded our collection by looking at the portfolios of the individual authors and/or data graphics team from which the exemplars were produced, and at newsrooms from underrepresented geographic regions in prior work on visual data journalism and narrative visualization. 
 Ultimately, we contacted 35 data journalist consultants, staff journalists, directors via email and/or social media who were credited as authors on these exemplars. Out of the 35 contacted, 17 (nine male, eight female) participated in our study: three consultants, ten staff journalists, and four directors from newsrooms based in different geographical regions, specifically nine in North America, one in South America, three in Europe, two in Asia, one in Oceania, and one in Africa. Six were cold contacts, nine were warm contacts within our personal network, and two were \new{referred by} other participants. 

\subsection{Interviews}
We conducted 17 one-on-one interviews, with each interview lasting between 30--60 minutes. We conducted the first of these interviews in person, and the subsequent 16 online over Zoom. With the exception of the first in-person interview conducted by the first author alone, all other participants were interviewed by a pair of authors~\cite{akbaba2023two}. The first author led the interview, while the second or last author took notes. 
We created and followed a semi-structured interview guide consisting of a series of questions that probed participants' beliefs about the design of a successful visual data story. Specifically, we asked participants to walk us through a story from their portfolio that best embodied their design philosophy. We then followed with questions regarding their treatment of datasets, visual and interactive design decisions, and perspectives on success and criticism of stories. More details about the interview guide can be found in supplemental materials.
We recorded the audio, video, and screen during each session to aid analysis. Following the session, we automatically generated a transcript of the audio recording for each the interview using the transcription functionality in Microsoft Word. The first author reviewed the generated file for transcription errors and anonymized the text. We anonymized participants according to their level of editorial agency in the organization: \textit{consultant}--CXX, \textit{staff journalist}--JXX, or \textit{director}--DXX. When reporting quotes from the interview corpus, we revised them to facilitate reading this paper, i.e., truncation, grammar, removal of filler words such as \textit{like}, \textit{you know}. Unedited quotes are available in supplemental materials.

\section{Interview Analysis}
\label{sec:theory} 

In constructing our study methodology, we planned an interview analysis strategy that acknowledges and enriches the analysis with the author team's experience as visualization researchers and practitioners. At the outset we considered a group diffractive reading~\cite{barad2018diffracting} with all coauthors of our interviewees' responses following the method outlined in Akbaba et al.~\cite{akbaba2023troubling}. Diffractive reading~\cite{barad2018diffracting} draws from feminist and interpretivist perspectives to support multiple perspectives and interpretations of the data. In essence, diffraction draws out differences, rather than similarities and patterns, within the data~\cite{akbaba2023troubling}. 
However, as our group discussions progressed, what became most salient was not our (the researchers') interpretative differences in analyzing the interview data, but rather the opposing beliefs about visual data journalism that hung in the silence within our interviews. We found ourselves asking how the \textbf{design philosophies and practices of the visual data journalists that we interviewed make visible tensions and contradictions in epistemic beliefs in visual data journalism}. 
We turned to deconstruction theory~\cite{derrida1974grammatology} to surface and disclose the opposing philosophical beliefs that define \textit{``successful''} reporting among the visual data journalists whom we interviewed. We employed genealogy~\cite{foucault2003essential} to understand the origins and shifting contexts of these beliefs as products of paradigm shifts associated with historical and sociopolitical events. 
\new{In the following, we introduce the core concepts and method for both deconstruction and genealogy that we have validated with scholars in comparative literature.}

 \begin{figure*}[ht]
  \centering
  \includegraphics[width=\linewidth, alt={Deconstructive analysis described in five parts, involving group analysis of points of tension on sticky notes, which feeds identifying a binary opposition, highlighting of interview quote excerpts that indicate implicit beliefs, and ending with collecting all quote analyses to collapse a binary opposition.}]{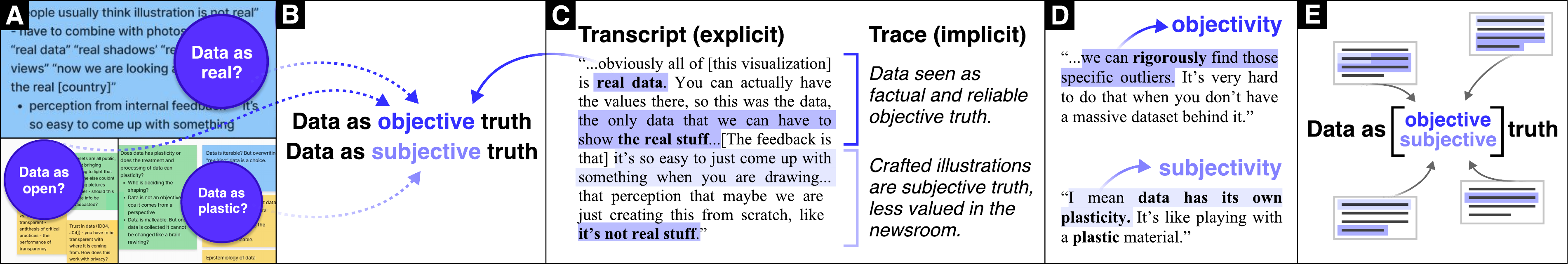}
  \caption{%
  	\new{Deconstructive analysis applied to the binary opposition \textit{objectivity/subjectivity}: After a close reading of the transcripts,  we A) identify points of tension (purple circles) in the interview corpus over multiple group analysis sessions on FigJam (provided in \href{https://osf.io/5fr48/}{supp. materials}), B) identify the binary opposition that best encapsulated this tension, C) identify the privileged binary term, i.e., objectivity, by analyzing the implicit beliefs (traces) within participant statements, D) locate additional statements in the interview corpus supporting or undermining the privileged term, and E) examine statements collectively to understand the collapse of this binary opposition. See \cref{sec:objective_subjective} for the full analysis.}%
  }
  \label{fig:analysis}
\end{figure*}

\subsection{Deconstruction theory}
Deconstruction is a poststructuralist theory best known through the work of Jacques Derrida that interrogates the relational quality of meaning in written or visual language~\cite{Brunette1994, derrida1974grammatology, derrida2004dissemination, rivkin2017literary, prasad2017crafting}. In the eyes of deconstruction, meaning is unstable---neither fixed nor absolute---because words derive meaning in relation to other words. By acknowledging that language is ``a will o' the wisp''~\cite{jameson2024years}, deconstructive reading affords expansive conversations about the embedded assumptions, contexts, and ideologies in literary works~\cite{tyson2023critical}, visuals~\cite{Brunette1994}, and interview transcripts~\cite{mazzei2004silent,jackson2011thinking}---in our case, our conversations with data journalists.

Core to deconstruction are the semiotic terms \textbf{signifiers} and \textbf{signified}, \new{previously explored in visualization rhetoric~\cite{hullman2011visualization}}. Signifiers are the words and symbols used to designate mental concepts, referred to as the signified. Deconstruction posits that a signifier points to multiple possible meanings, which connect then to other possible meanings. As an example, a \textit{`red traffic light'} can signify not only a mental image of stopping, the colour red, or a physical traffic light, but also mental concepts of danger, roads, cars, and traffic laws. 

A second concept in deconstruction is the notion of a \textbf{trace}, that is, what is left unexpressed but \new{implicitly} impacts meaning. Continuing our example, the absence of a green light signifying \textit{`go'} \new{influences} the meaning of \textit{`red traffic light'} as \textit{`stop.'} \new{In deconstruction terms, the absent green light is a trace for the concept \textit{`stop.'}} 

Central to deconstruction is the concept of \textbf{différance}. Meaning both \textit{difference} and an \textit{act of deferring meaning}, différance refers to the fluid underlying assumptions that allow a signifier to take on different meanings through changing circumstances~\cite{tyson2023critical}. \new{Returning to our example,} a \textit{`red traffic light'} is meaningful because contemporary society differentiates \textit{`red'} signifying \textit{`stopping'} from \textit{`green'} signifying \textit{`going.'} The underlying assumption is that we acknowledge and obey traffic laws \new{only where} \textit{`red'} means \textit{`stop.'}

Through these concepts, deconstruction searches for \textbf{implicit contradictions} in language. In Western philosophy, these contradictions are binary and hierarchical, with a pair of terms in which one term is \new{presented as superior, i.e.,} privileged, in \new{opposition} to the other, e.g., \textit{nature/culture}, \textit{good/evil}, \textit{reason/emotion}, \textit{mind/body}. Identifying this privileged term affords insights into the \new{beliefs} of a group or society. Deconstructing a \textbf{binary opposition} does not aim to invert or destroy this hierarchy, but instead to \textbf{show where these terms share commonalities}. If we consider \textit{`red traffic light'} to rest on the binary terms \textit{safety/danger}, we can say that society privileges safety, yet these terms overlap: one does not encounter a red light unless driving in traffic, which is a statistically dangerous activity. The fact that traffic lights exist to keep people \textit{safe} is a simultaneous acknowledgement of \textit{danger}.

\subsection{Deconstructive analysis of data journalist interviews}

Our deconstructive analysis of interview transcripts is motivated by Jackson \& Mazzei~\cite{jackson2011thinking} and Douglas~\cite{douglas2017beyond}, who argue that traditional interpretive or constructivist methods, e.g., thematic analysis~\cite{braunthematic2021}, are limited in constructing themes and patterns by what is said or shown. Through poststructural reading, e.g., deconstruction, the researcher is asked to investigate both what is said and unsaid, thus reframing research questions in larger contexts of norms and beliefs. 

We conducted our analysis following the group analysis protocol outlined Akbaba et al.~\cite{akbaba2023troubling}, with the exception that we read the transcripts through a deconstructive, rather than diffractive, lens. We divided the analysis work over the course of five sessions lasting two hours on average. Each session focused on three to four interview transcripts grouped by participant level of editorial agency in the organization, i.e., \textit{consultant}, \textit{staff journalist}, or \textit{director}. For each session, all coauthors individually read the interview transcripts and recordings before gathering to discuss salient aspects of the interview corpus. Sessions focused on identifying points of tension within the interview corpus, typically when participants expressed differing or contradicting views within and across interviews. The first author used an online whiteboarding tool, Figjam\footnote{https://www.figma.com/figjam/}, to memo throughout the session. Per Tyson~\cite{tyson2023critical}, our deconstructive reading of participant interview transcripts took the following sequence \new{(for a detailed example of our workflow see \cref{fig:analysis})}. Questions probing for \textbf{signifier(s)}, \textbf{trace}, and \textbf{différance} drew from prior work on deconstructive reading of interview transcripts~\cite{mazzei2004silent, jackson2011thinking, douglas2017beyond}. Our goal throughout was to answer the question: \textit{How do the design philosophies and practices of visual data journalists make visible tensions and contradictions regarding epistemic beliefs in visual data journalism?} 
 
\begin{enumerate}[nosep]
    \item Identify a central tension (expressed through a \textbf{signifier}/chain of signifiers) in the interview corpus \new{(\cref{fig:analysis}A)}. 
    Here, we ask, \textit{what words (signifiers) describe this tension, and what is the tension intended to represent (the signified)?}
    \item Find the \textbf{binary opposition} that this tension rests on \new{(\cref{fig:analysis}B)}. 
    \item Identify which of the binary terms is privileged. The ``side'' of the opposition that the speaker supports reveals the implicit belief (the \textbf{trace}, \new{\cref{fig:analysis}C)}. We ask, \textit{what words or concepts are left unsaid to maintain a self-contained and self-sufficient truth?}
    \item Locate statements in the transcripts that conflict with or undermine this privileged hierarchy (cases of \textbf{différance},  \new{\cref{fig:analysis}D)}. We ask, \textit{what underlying tensions, assumptions, and biases arise from the use of an unstable signifier? What is the distance between the explicit and implicit meanings of an unstable signifier?}
    \item Show how the hierarchy collapses through our analysis \new{(\cref{fig:analysis}E)}.
\end{enumerate}

\new{After close readings of the transcripts, the authors identified the points of tension, binary oppositions, and privileged terms through group discussions. The first author pulled participant statements that reflected each side of the opposition; these selections were validated by the last author. Finally, the first and last author examined the statements collectively in~\cref{sec:deconstruction} to show a collapse of the binary oppositions.}

\subsection{Genealogical analysis of binary oppositions}

Deconstruction taken to its extreme can lead the reader to a place where there is no meaning~\cite{tyson2023critical, jameson2024years}. This is not only frustrating, but unproductive. These binary terms are unstable in the present as well as through time and across different cultures~\cite{tyson2023critical}. 
We adopted Foucault's genealogy~\cite{tamboukou1999writing, foucault2003essential, koopman2013genealogy} which threads the history of objects, rules, or ideas that may be otherwise considered ahistorical to show their branching, often contradictory past. This technique contextualizes the historical changes and conditions---social, technical, and political---to provide a holistic understanding of beliefs on \textit{data as objective/subjective truth} and \textit{design as humanistic/mechanistic insight} \new{as they impact data journalists' professional practices.}
This approach extends our critique of visual data journalism perspectives by exploring the variations in différance, the movement of meaning, over time, following in the vein of Akbaba et al.~\cite{akbaba2024entanglements}'s genealogy of entanglement and Correll \& Garrison`s~\cite{correll2024body} genealogy of data cultures in medical illustration and visualization. 

To thread our observed binary oppositions through time, we set our historical starting point as the time period when any one of the terms within our binary oppositions first emerged. Advancing in time from the historical starting point, we identified the major philosophical positions associated with the modern definitions of our binary oppositions (per Sec.~\ref{sec:terms}), e.g., \textit{empiricism} with \textit{objectivity}~\cite{bach2018narrative} or \textit{Descartes' mind--body dualism} with \textit{mechanism}~\cite{descartes2008meditations}. We looked at paradigm shifts~\cite{kuhn1997structure} and their concurrent historical and sociopolitical contexts that affected prominence or waning of these positions. Within each paradigm shift, we localized representative examples of visualizations we saw as possibly influencing the perspectives of our participants, \new{e.g., Playfair's statistical graphics}, which reflect the then-privileged epistemic virtue of the period\new{---in this example, \textit{objectivity} through \textit{empiricism}~\cite{playfair1802elements,russell2004history}}. 
Since our study centres on visual data journalism, which has strong roots in statistical graphics, we prioritized sources examining the philosophy and history of visual scientific knowledge~\cite{robin1992scientific, kuhn1997structure, daston2021objectivity, drucker2014graphesis} extending into statistical visualization and data-driven journalism~\cite{friendly2001milestones, friendly2008brief,howard2014art, rendgen2019history}.
We acknowledge the anachronism of applying these modern definitions to other historical periods. This \textit{methodological anachronism} allows for characterization of objects and events of the past without taking sides in their debates~\cite{schneider2014anachronism}, and to disclose tacit shifts or repetitions within other contexts that are otherwise less easily discernible~\cite{shiner1982reading}. 

\section{Results I: Deconstruction of différance}
\label{sec:deconstruction}

We focus in this manuscript on the emergence of two binary oppositions for the visual design of data in newsrooms: 
1) \textit{objectivity/subjectivity} in the value and interpretation of data,  
2) \textit{humanism/mechanism} in designing for insight. 
In this section, we show how our deconstruction approach revealed these binary oppositions and exposed the explicit and implicit ideologies underlying visual data journalism. \new{Within each quote, we colour-code text by their associated binary term. The darker colour indicates the privileged term as seen across all interviews. \textbf{Signifiers} are bolded while \textit{traces} are italicized. }

\subsection{Data as \textit{objective/subjective} truth}
\label{sec:objective_subjective}

\begin{figure}[h]
  \centering
  \includegraphics[width=\linewidth, alt={Deconstructive analysis for objectivity and subjectivity involves looking at a chain of signifers leading from data to truth and their traces.}]{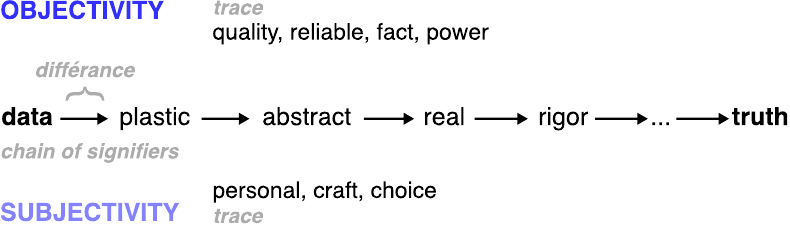}
  \caption{%
  	A visual overview of our deconstructive analysis of interview transcripts for objective/subjective beliefs in \textit{data as truth}.%
  }
  \label{fig:objective-subjective}
\end{figure}

We find that participants perceive data to signify many things, some of which are stated explicitly \new{(signifier)} while others implicitly \new{(trace)}. ~\cref{fig:objective-subjective} \new{summarizes the chain of explicit signifiers and implicit trace(s) that we observed to support a given perspective, e.g., the unstated belief that objectivity yields high-quality and reliable data stories}. Our participants prioritize principles of objectivity through their references to data as \textbf{real} or \textbf{rigorous}, even though their treatment of data in the crafting of a successful story reflects subjectivity, e.g., data as \textbf{plastic} or \textbf{abstract}. \new{In the following, quote excerpts reflecting a \obj{privileging of objectivity} are highlighted in a darker blue relative to excerpts reflecting a \subj{privileging of subjectivity}}.

We first encountered a subjective view of data in conversation with J07. Tabular data in this case was perceived as \textbf{plastic}:
\vspace{-0.15cm}
\begin{quote}
\label{J07plastic}
    ``\subj{I mean data has its own \textbf{plasticity}. It's like playing with a \textbf{plastic} material.} When I'm working with data, I always start by [...] visualizing it, and then \subj{changing stuff and making it, you know, kind of sculpting it.}'' --J07
\end{quote}
\vspace{-0.15cm}
This notion of data as \textbf{plastic} affords \textit{flexibility} in storytelling. While our journalists use data to expose the \textbf{realities} of a situation, they acknowledge that data is an \textbf{abstraction} of human experiences that readers can find difficult to connect with. Our conversations with D01 and J08 best illustrate the strategies journalists use for large, hard-to-conceive quantities, e.g., choosing a single data point as a hook for the story. D01 takes a \textit{personal} approach by using the audiences' own data, while J08 describes how outliers can drive the message:
\vspace{-0.15cm}
\begin{quote}
    \label{D01me}  
    ``\subj{We try to put a reader at the centre of the data,} but also try to make the data feel a little bit \obj{more visual, a little bit more graspable, touchable, physical} [...] I like to call [this approach] the \subj{\textit{`me factor'}, or the \textit{`me layer.'}} How do I compare to the rest of the population? \obj{That makes data feel less \textbf{abstract}.}'' --D01

    \label{J08outlier}
    ``\subj{We [journalists] are attracted to those extremes, in part because they're dramatic, in part because that's what grabs people's attention and underscor[es] the point} [...] It's important when you're doing any kind of data journalism to be grounding the concepts and the stories and the takeaways in \obj{\textbf{real} physical things} in the \obj{\textbf{real} physical world} [...] \obj{this is a \textbf{real} brick and mortar place, \textbf{real} flesh and blood.} Humans live here.'' --J08

\end{quote}
\vspace{-0.15cm}    
While journalists valued and drew upon various forms of subjectivity in their practice such as narratives and creative expertise, there is an \textit{unspoken standard, i.e., a trace, that data is a prerequisite for a successful, legitimate story}. The presence of data, especially data that is numerical, quantitative, and measurable, adds a certain \textbf{reality}, \textit{strength} and \textit{factuality} to their reporting that is, at times, at odds with creative expertise in traditional newsrooms. In conversation with J09, we noticed a dissonance between their pride as a trained illustrator and skepticism towards illustrations in the newsroom. In particular, we saw the signifer \textbf{real} being used frequently to describe certain forms of visual information, particularly video footage, photographs, and 3D spatial data visualization.
\vspace{-0.15cm}
\begin{quote}
\label{J09real}
    ``[...] obviously all of [this map visualization] is \obj{\textbf{real data}.} You can actually have the values there, so \obj{this was the data, the only data that we can have to show the \textbf{real} stuff.}'' --J09
\end{quote}
\vspace{-0.15cm}
 In juxtaposition, the absence or even negation of the word \textbf{real} is used with \textit{hand-crafted} illustrations that were drawn over screenshots of a spatial visualization. When asked about this tension between what visualizations are considered \textbf{real} and \textbf{not real}, J09 elaborates on how their internal editorial team perceives illustrations:
\vspace{-0.15cm}
\begin{quote}
    ``\subj{[... the feedback is that] it's so easy to just come up with something when you are drawing} [...]
    that maybe we are just creating this from scratch, like \obj{it's not \textbf{real} stuff.}'' --J09
\end{quote}
\vspace{-0.15cm}
We see further this unexpressed meaning of data as \textit{reliability} and \textit{quality} through J08, who expresses that anecdotes and outliers can only be \textit{responsibly} and \textbf{rigorously} used if the dataset behind a story has statistical \textit{strength}. J08 takes pride in their data story on social injustice following ``\obj{an almost academic or scientific method}'' and attributes their responsibility to \textbf{rigour} as being ``beholden to editorial ethics.''
\vspace{-0.15cm}
\begin{quote}
\label{J08rigor}
    ``\obj{It's not enough to cherry pick a single, weird anecdote and try to write a story off of it. But when you have the the weight of the investigation behind [an anecdote],} \subj{you can use that singular entry point to get people invested in the story.} [...] It's a tool in the toolkit that data journalism can do well because \obj{we can \textbf{rigorously} find those specific outliers. It's very hard to do that when you don't have a massive dataset behind it.}'' --J08
\end{quote}
\vspace{-0.15cm}

Other participants similarly elude to the \textit{strength} and \textit{power} of data with respect to political activism, where stories built on data can be used to empower readers for action. J10 expresses how visualization can call attention to power imbalance between government and citizen. 

\vspace{-0.15cm}
\begin{quote}
\label{J10power}
    ``We thought this would be a great time to think about how people in government positions can have a budget of \$10 billion, whilst the continent is struggling to find [electrical] power. [...] \obj{We wanted an easy way for people to have an \textbf{understanding} of just how much \$10 billion is.} [...] If you're seeing this [chart], if we had \$10 billion, this is what we could do and this is how we would power the continent.'' --J10
\end{quote}
\vspace{-0.15cm}
D04 mentions their country's ``strong coercive laws around news'' and views the gathering of social statistics as activism:
\vspace{-0.15cm}
\begin{quote}
\label{D04activism}
    ``Sensitive data around race, incarceration, sexual violence [...] are \obj{not publicly \textbf{available}, so you have to find your own way to gather [them]}. This is the data activism side of things [...] \obj{[We've] manually gathered our own data [by] working with communities} [...] particularly with migrant workers and also sexual violence. Community has become a solution to mitigating that and making sure that \obj{every data point has a source linked to it.}'' --D04
\end{quote}
\vspace{-0.15cm}
J04, who works in the same country as D04, echoes this sentiment towards data activism. They recognize the absence of data as a limitation: 
\vspace{-0.5cm}
\begin{quote}
    ``\obj{We are \textbf{missing} data [...] It's very hard for us to do a story that's entirely driven by data} [...] I don't think it should be that way though. [...] \obj{that's the limitation that we have to work with.}'' --J04
\end{quote}
\vspace{-0.15cm}
Through contradictory \new{sentiments}---cases of \textit{différance}---found \new{within and across participants}, we interpret the need and reverence for data as a privileging of \textit{data as objective truth}. Even though participants embrace and take pride in their \new{subjective framing of} data, e.g., \textit{personalizing, crafting, and choosing} how data plays a role in their story, they share an implicit belief that \textit{objectivity---in the form of data---is required for successful journalistic reporting}. Abiding by this belief allows participants to position data-driven reporting as \textit{high-quality, reliable, factual, and powerful}. When data is \textbf{missing} and unavailable, objectivity quickly becomes a limitation in the stories that can be told. 

We see a collapse of the binary opposition in which \textbf{subjectivity in reporting \new{is contingent upon the} objective truth that data is seen to bring to storytelling}. \new{As captured by J08, one can only use a ``single, weird anecdote'' when there is a ``massive dataset behind it.''}

\subsection{Design for \textit{humanistic/mechanistic} insight}

\begin{figure}[h]
  \centering
  \includegraphics[width=\linewidth, alt={Deconstructive analysis for humanism and mechanism involves looking at a chain of signifers leading from design to insight and their traces}]{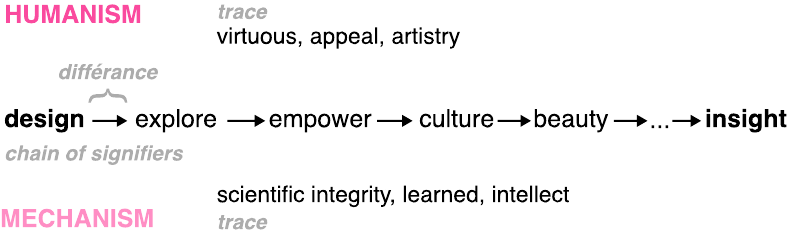}
  \caption{%
  	A visual overview of our deconstructive analysis of interview transcripts for humanistic/mechanistic beliefs in \textit{design for insight}.%
  }
  \label{fig:humanism-mechanism}
\end{figure}

Participants express two views on the \textbf{insights} that visualization design reveals, as summarized in~\cref{fig:humanism-mechanism}. While visual data journalists acknowledge the mechanistic insight design affords e.g., strategies to capture \textbf{attention} or build a\textbf{ mental model} of the dataset, they make careful, virtue-driven choices when employing these strategies to depict the emotional \textbf{weight} of data and appeal to the \textbf{humanity} of the audience. \new{In reporting quotes, we again use a darker colour to highlight a \hum{privileging of humanism} relative to excerpts reflecting a \mech{privileging of mechanism}.}

A commonality among the journalists is the idea that design, in particular interactivity, lets someone satisfy their curiosity and, per D02,  ``[get] insight from actually \mech{actively \textbf{exploring} things}''. Interactivity for \textit{intellectual} insight, specifically \textbf{mental model} building, is best illustrated in conversation with D02: 
\vspace{-0.15cm}
\begin{quote}
\label{D02construct}
    ``[...] the role for us as tool builders is \mech{what is the fastest way to help a researcher get a \textbf{mental model} of what's actually in that dataset}? [...] \hum{[I] was always really struck by constructivism and Piaget’s work with this idea [that] you don't just sit back and get information thrown at you.} [...] I think that's the core of what you're doing when you're building interactive things, right? \mech{How do I take this dataset and, instead of making the static figure, making a thing that lets you play the data back [...] poke at that dataset and mess with it.}'' --D02
\end{quote}
\vspace{-0.15cm}

This constructivist view is also echoed by J02, who uses scrolling interaction, and C02, who uses a slideshow interaction, to layer information in order to build their readers' \textbf{mental model}. Interactivity is also seen as \textbf{empowering} by J08 in that it affords more granular \textbf{exploration} of a dataset. This is echoed by D04 who expresses that ``\mech{granularity is important} for the story in terms of [getting] further insight [in]to the dataset.'' 
\vspace{-0.15cm}
\begin{quote}
\label{J08map}
    ``[...] the goal of [an \textbf{exploratory} map] is to try to \hum{\textbf{empower} journalists or researchers or just people who are curious about what their city looks like.} [...] \mech{it's much more granular and it's much more designed around giving people access to [the data].}'' --J08
\end{quote}
\vspace{-0.15cm}

\new{Publicly-available datasets with sufficient granularity can be a gold mine for data journalists. While D04 and J04 struggle in their countries to gather data, J08 describes a dataset triangulated from several open datasets in balance with privacy and insight: ``somebody obviously could do the individual leg work'' to find personal information behind the data, ``\mech{just as how you could do that if you had any address}.''}

While most participants acknowledged the value of design principles for \textit{cognitive clarity}, we find an underlying consideration and care for the humanistic aspect of mechanistic design principles. Alongside a curiosity for statistical insights is a curiosity about the people from whom data is collected. As J02 expresses, ``\hum{it is immediately pretty tantalizing to follow one person throughout a day.} I wonder what else [this person is] going to do.'' Some participants experiment with how mechanistic strategies can translate humanistically. J02 discusses the use of motion for ``\hum{injecting life}'' into data \new{to \textbf{empower} the people behind the data points}, when mechanistically motion is used for ``\mech{getting \textbf{attention} to the right parts of the story}.''
\vspace{-0.15cm}
\begin{quote}
\label{J02motion}
    ``It's really important when we visualize information about people, \hum{especially people who are often not given the dignity.} [...] It was important they move because \hum{it's important you see everyone on here as \textbf{human} or as a representation of a \textbf{human}.}'' --J02
\end{quote}
\vspace{-0.15cm}

J07 similarly discusses humanistic considerations behind their interaction design choices for a memorial piece, specifically each victim being equally represented and included in the design. 
\vspace{-0.15cm}
\begin{quote}
\label{J07law}
    ``There is this passage of [my country’s] law that says if two people died in the same event and their exact time of death is not well known, it is said that no one died before the other. \hum{We have to show everyone at the same moment.} [...] \hum{[Individuals are] all going to have their place in this glyph} [...]
    but they're not going to come in or out of it. \hum{We're not using a filter interaction.}'' 
    --J07
\end{quote}
\vspace{-0.15cm}
Some participants go as far to disapprove of mechanistic principles when depicting the \textit{emotional or ethical} \textbf{weight} of data, particularly for heavy and sensitive stories, e.g., on human tragedy. J01 expresses how statistical design principles cannot capture the ``\textbf{weight}'' of lives lost; instead, visual metaphors can ``[present] statistics in ways that make you \hum{understand what is behind that number}.'' 
\vspace{-0.15cm}
\begin{quote}
\label{J01weight}
    ``How do you get people to understand that [...] \hum{each number has a \textbf{human} story behind it}? [...] How do you communicate this without showing them as \mech{[simply] circles on a screen or dots in an animated bubble [chart]}? Those types of approaches felt very \hum{sterile and almost not appropriate to tell the data which carries so much \textbf{weight}}.'' --J01 
\end{quote}
\vspace{-0.15cm}
J07 also comments on the mechanistic connotations of the words ``data visualization.'' There is also an expectation that data visualizations come with a legend or ``\textbf{instructions}'' on how to read it, which ``\hum{takes some of the magic out of}'' an \textit{emotional experience}.
\vspace{-0.15cm}
\begin{quote}
\label{J07instruct}
``I definitely didn't want [instructions for] the memorial because \hum{no memorial comes with an \textbf{instruction} set.} [...] I wouldn't even call this particular piece, this memorial, data visualization [...] \hum{It’s not respectful to say that this was data visualization.}'' 
--J07
\end{quote}
\vspace{-0.15cm}
Design brings out both the humanity behind the data points and the humanity of the journalist behind the reporting. J05 expresses leaving a mark of themselves in their work.
\vspace{-0.15cm}
\begin{quote}
\label{J05handdrawn}
    ``I touched up on the edges [of the chart] so it looks like there's a little bit of edging and hand-drawn qualities [...] I think it's a good way to let people realize that \hum{this work has hands behind it.}'' --J05 
\end{quote} 
\vspace{-0.15cm}
D04 echoes this sentiment on a larger scale, sharing how \textit{aesthetics} allows their newsroom to bring out the \textbf{culture} behind the data, the reporting, and the audience:
\vspace{-0.15cm}
\begin{quote}
\label{D04culture}
    ``We taught ourselves to be an anti-Orientalism, non-Eurocentric [outlet], which means that \hum{even in our visual \textbf{culture}, all the vernacular that we use in our visual design, we try to reference indigenous local motifs and designs} and thinking around how we present this information. 
    [...] \hum{When we talk about shared \textbf{culture} within Asia, we often use a perspective of delight and pride,} and that's intentional because \hum{we want to give that community
    that sense of representation that is not from this mystified lens.}
    '' --D04
\end{quote}
\vspace{-0.15cm}
However, appreciation of \textit{aesthetics} is context-dependent.  Journalists need to, as per D02, ``able to speak in a language that \hum{[the audience] are comfortable with} and \mech{makes sense and is understandable}.'' When designing for a scientifically-trained audience, D02 laments that a ``\textbf{beautiful}'' design can create doubts about the \textit{scientific integrity} of the visualization. The latter perspective on \textbf{skepticism} towards overly \textit{aesthetic design} is reflected in the works of science journalists J03 and D03. They retain the visual idioms and chart types when redesigning statistical figures taken from academic publication, implicitly following mechanistic design principles so as to support fluent understanding by the scientific community. 
\vspace{-0.15cm}
\begin{quote}
\label{D02beauty}
    ``\mech{We need to be able to build all of that infrastructure so that we can pull in all that data and make use of it,}  and then \hum{design wise make [data] actually \textbf{palatable} for humans.} [...] if the design was getting a little too refined or too fancy, [scientists] responded to it with more \textbf{skepticism} [...] \mech{it's so \textbf{beautiful}, it doesn't feel like science.}'' --D02
\end{quote} 
\vspace{-0.15cm}

Through cases of \textit{différance} in our interview corpus, we interpret the underlying desire to use \textit{design to appeal to the humanity of the audience} e.g., designing for the \textit{emotional} \textbf{weight} of data or its \textbf{palatability}, as privileging humanistic perspectives towards visualization design. However, participants \textbf{cannot use the language of design without its mechanistic connotations and theoretical foundations}, e.g., design principles derived from behavioural and cognitive studies. Though our participants prioritized humanistic perspectives on design, we see the interdependence of humanistic intentions on traditionally mechanistic principles as a collapse of this binary opposition.

\section{Results II: Genealogy of visual data and insight}
\label{sec:genealogy}

\new{Through deconstructive analysis, we surfaced the privileging of certain beliefs while showing the difficult and sometimes subtle contradictions between what data journalists express and implicitly believe about their work. The objects, rules, and ideas that shape participants' beliefs of \textit{data as objective/subjective truth} and \textit{design for humanistic/mechanistic insight} are not self-enclosed but rather rooted in different, sometimes multiple, points of our history. Identifying these roots through genealogical anaylsis~\cite{foucault2003essential} provides essential social, technical, and political context to unpack contemporary journalists' beliefs on success in their work.} Our analysis is scoped to events in Western history as we, the coauthors, observed Western visualization design philosophy exerting an undeniable influence on participants' design practices, regardless of where their newsroom was located. 

\begin{figure*}[ht]
  \centering
  \includegraphics[width=\linewidth, alt={A timeline tracing the privileging of the two sets of ideologies, humanism and mechanism as well as objectivity and subjectivity. Above the timeline is a series of historical and contemporary visualizations representative of the privileged ideologies at the time.}]{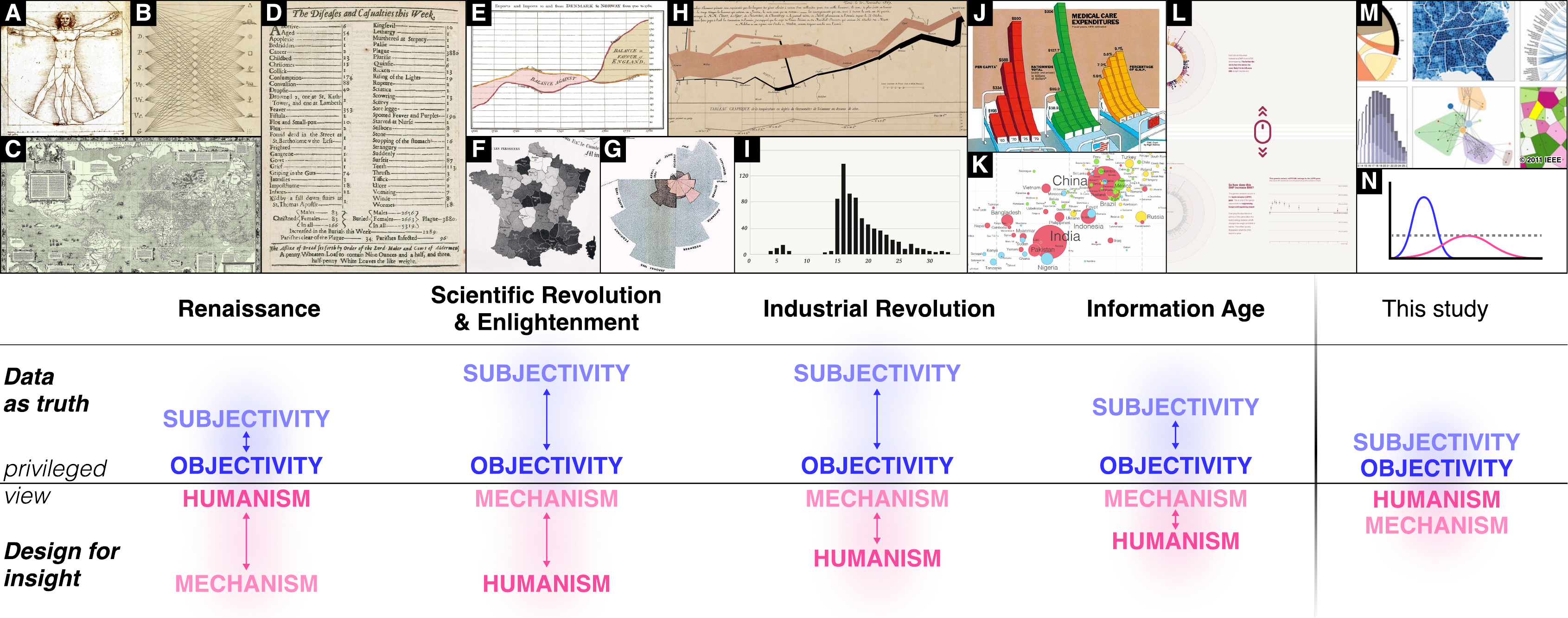}
  \caption{%
  	A timeline tracing the fluid and intertwined lineages of \textit{objectivity/subjectivity }and \textit{humanism/mechanism} from the Renaissance to Information Age, relative to our current study. Alongside this timeline, we position data visualizations representative of the privileged ideologies at the time. While there is a longstanding view of data as objective truth and design for mechanistic insight throughout the course of data visualization history, we observe a collapse of these opposing beliefs through a ``renaissance'' and embrace of subjectivity and humanism in today's visual data journalism.%
  }
  \label{fig:genealogy}
\end{figure*}

\new{In this section, we explore the historical lineage of the key tensions we found through deconstruction: \textbf{(1) truth is visual and quantifiable through data that are subjectively framed}, and \textbf{(2) ``serious'' design should be detached from human influence, yet human intervention creates insight.}} Our earliest point of reference is the Renaissance, chosen as this period marks the emergence of one of our binary terms, \textit{humanism}, as a theory~\cite{russell2004history}. \new{We observe différance as rooted in times of major paradigm shifts~\cite{kuhn1997structure}:} the Scientific Revolution and Enlightenment (we group these periods together, since we found similar privileging of binary terms between the two periods), Industrial Revolution, and the Information Age. \cref{fig:genealogy} visually summarizes the fluid, intertwined lineages \textit{objectivity/subjectivity} and \textit{humanism/mechanism} that underpin the (un)spoken beliefs of our participants. \new{We find that while objectivity remains the explicitly privileged view through history, humanism and mechanism have interchanged. In the following, we emphasize in \textit{\textbf{bold}} the beliefs that we surfaced through our deconstructive analysis and support them with quotes from the interview corpus.}

\new{\textbf{Truth is visual and quantifiable through data that are subjectively framed---}}%
\new{Our participants' \textbf{\textit{curiosity about the human condition}}, such as J01's desire to tell the ``human story behind [data]'', can be traced back to  Renaissance \textit{humanism}~\cite{russell2004history}. Many participants accordingly \textbf{\textit{believe truth to come from sensory experience and observations}}, reflected in D01's preference to ``make the data feel a little bit more visual, a little bit more graspable, touchable, physical.'' Participants leverage \textbf{\textit{vision}} in particular, creating graphs and charts as ``an easy way for people to have an understanding [of data]'', as expressed by J10. These views are rooted in the centuries-old traditions of} Renaissance scholars who aimed to uncover and represent hidden truths about natural phenomena through a scientific approach grounded in \textit{empiricism}~\cite{locke1847essay, bacon1910advancement}, which holds that universal truth and insight can be extracted from observations by the naked eye. The prioritization of sensory experiences to find truth, with vision at the top of the hierarchy~\cite{robin1992scientific}, forms a through-line into participant beliefs today.

\new{Participants view \textbf{\textit{truth as quantifiable to varying degrees}}, most explicitly captured in J09's comment that quantitative values are ``real data'' that can ``show the real stuff.'' J08 reflects that a ``massive dataset'' adds weight to an investigation and allows stories to be told ``rigorously'' through a more ``academic or scientific method.'' When data are missing or not publicly available, we recall that D04's team ``manually gather their own data'' to expose the truth of a situation. These beliefs in data as a quantifiable truth are the legacy of} major paradigm shifts during the Scientific Revolution and Enlightenment that include experimentation, the Copernican revolution, and Newtonian mechanics~\cite{gillispie2016edge}, \new{all of which pushed for a mathematical, ``objective'' understanding of the natural world~\cite{kuhn1997structure}. This view remains explicitly privileged today} under the guise of data reverence and \textit{positivism}, a position holding that all genuine knowledge is either true by definition or derived by inductive reasoning and logic from sensory experience~\cite{gillispie2016edge}. 

\new{\textbf{\textit{Yet observed, quantified data may not convey the fundamental nature of the entity}}. While our participants believe empirical observations to be real and factual, i.e., a privileging of scientific objectivity, we simultaneously see that they may \textbf{\textit{render an idealized version of these observations, believing visualizations to be capable of closing the distance between quantitative data and truth}}. Participants craft and choose the ways they represent \textit{data as truth}, such as J08's use of outliers as a ``singular entry point to get people invested in the story'', or D01's strategy of ``put[ting] the reader at the centre of the data.'' This subjective treatment of data echoes Renaissance through to Enlightenment scholars' \textit{truth-to-nature} ideals in pursuit of \textit{objectivity}. These ideals led to the production of visualizations that venerated observed forms and structures, e.g., idealized proportions of the human body (\cref{fig:genealogy}A), human ways of knowing (\cref{fig:genealogy}B), and geographical knowledge corrected with Mercator projection (\cref{fig:genealogy}C).}
\new{This belief in the power of visualization for quantifiable truth is reflected through historical statistical graphics (\cref{fig:genealogy}D--H)} that aimed to draw mental correlations between what seems at times disparate information. Playfair, who invented numerous types of statistical graphics to communicate relationships regarding economics and societal progress, wrote that ``the best way to capture the imagination is to speak to the eyes'' as they were ``the best judge of proportion, being able to estimate it with more quickness and accuracy than any other of our organs~\cite{playfair1802elements}.'' 

\new{\textbf{``Serious'' design should be detached from human influence, yet human intervention creates insight---}}%
\new{Participants express \textbf{\textit{sentiments that undermine their creative expertise}}, e.g., D02's visualizations are seen as ``so beautiful'' that they do not ``feel like science'', or J09's illustrations perceived as ``not real stuff'' as it is ``easy to come up with something while drawing.''} This valuing of data over expertise is reflected through \textit{mechanical objectivity}~\cite{daston2021objectivity}, an epistemic virtue stemming from the Industrial Revolution that emphasized the removal of subjective interpretation from observations recorded by mechanical instruments and similarly sought to mechanize human productivity~\cite{russell2004history}. Rather than idealizing what they observed, scholars detached themselves from the final visualization so as to minimize human influence and “let nature speak for itself”~\cite{daston2021objectivity}. Mechanization can be traced further back to the Scientific Revolution. Descartes'~\cite{descartes2008meditations} ideas of separating the mind and body---where the mind governed conscious thought while the body was governed by physical laws---set the stage for a mechanistic view of nature, where nature was represented as a complex yet quantifiable machine~\cite{hooke1665micrographia} while the human mind remained apart. Separating mind from body saw to the separation of the qualitative from quantitative, experiential from measurable, and importantly the humanistic context of data, e.g., cultural, emotional, from the mechanistic, e.g., statistical, numerical~\cite{daston2021objectivity,russell2004history}. 

\new{However, participants believe that a \textbf{\textit{serious and rigorous visualization requires human intervention and expertise to facilitate insights and engagement with data}}. Journalists' goals can be productive and mechanistic to afford insights for their audience as efficiently as possible, e.g., D02 designing the ``fastest way'' to construct a ``mental model'' or J02 ``getting attention to the right parts of the story.'' Yet goals can also be personal and humanistic, e.g., J02 uses motion to ``inject life'' into data, while D04's team references indigenous local motifs to represent community rather than through a ``mystified lens.'' These creative differences in treatment of data and design resonates with} \textit{trained judgment}~\cite{daston2021objectivity}---a scientific attitude towards expertise that interprets the raw data of mechanical objectivity. 
 Within these creative differences lie tensions between mechanistic and humanistic design principles, not unlike the differing approaches between Edward Tufte---who famously advocated for a minimalist, statistics-driven approach of ``graphical excellence''~\cite{tufte1983visual} (\cref{fig:genealogy}I)---and \textit{Time}’s Nigel Holmes---who advocated for a illustrative, playful, and narrative approach~\cite{rendgen2019history} (\cref{fig:genealogy}J). 

\new{Despite the benefits to newsroom productivity, a number of participants believe that \textbf{\textit{statistics-driven approaches that draw from mechanistic ideals are ``sterile''}}, per J01, ``and almost not appropriate to tell the data which carries so much weight.'' 
This sentiment references criticisms towards mechanistic reduction of human productivity~\cite{maslow1954} during the paradigm shift from the Industrial Revolution to the Information Age.}  During this time, postmodernists such as Kuhn~\cite{kuhn1997structure} and the emergence of science and technology studies (STS) emphasized the relativity and subjectivity of data along with its inseparability from human perspectives, e.g., historical, social, cultural, ethical contexts. As data became synonymous with transmissible computer information, journalism benefited from the productivity that statistical methods and exploratory data analysis~\cite{tukey1977exploratory} brought to the newsroom, e.g., sifting enormous amounts of data and documents in the \textit{Wikileaks Afghanistan War} files~\cite{guardian2011wikileaks}. The rise of web interactivity (\cref{fig:genealogy}K--M) motivates our participants to further experiment with the ways that they engage their readership with data. Through interaction, they may explore the subjective, personalized experience that a data story can bring to an audience as well as the emotions multimedia can evoke. The ubiquity of visualizations in digital media, e.g., \textit{Flatten the Curve} (\cref{fig:genealogy}N), cultivates data literacy in the public. 

\new{Within the context of our participants during the Information Age, \textit{objectivity/subjectivity} and \textit{humanism/mechanism} lie in close tension. These apparent tensions are shaped by sociocultural, technical, and political rules and ideas, e.g., truth-to-nature, data as transmissible computer information, and social statistics to measure human productivity. These are simultaneously contemporary and historical concerns which mould the beliefs journalists hold today to assess their work.}

\section{Discussion}
\begin{quote}

``Graphical tools are a kind of \textbf{intellectual Trojan horse}, a vehicle through which assumptions are cloaked in a rhetoric taken wholesale from the techniques of the empirical sciences that conceals their epistemological biases under a guise of familiarity~\cite{drucker2014graphesis}.''
\end{quote}
\vspace{-2mm}
Our genealogical analysis showed the collapse of the binary tensions \textit{objectivity/subjectivity} and \textit{humanism/mechanism} to be a result of historically fluid epistemic beliefs about data visualization. We discuss these implications in visual data journalism~\new{and visualization at large}.

\subsection{\new{Implications for visual data journalism}}

Through deconstructive and genealogical analysis, we see the extent in which the binary terms that we identified overlap: subjective approaches to reporting cannot be embraced without historical views of data as objective truth; humanistic design approaches cannot be articulated without using design language and concepts with historically mechanistic connotations. This collapse is consistent across all participants regardless of their position in the newsroom. 

Objectivity has been posited as a strategic ritual~\cite{tuchman1972objectivity} that protects newsmen from risks of the trade, e.g., criticism from supervisors, deadline pressures, libel suits, and ultimately sustains the productivity of the newsroom. In our interviews, we observe this strategic ritual through the perceived reliability, quality, factuality, and power that data brings to the newsroom. By abiding by this ritual and these longstanding views of \textit{data as truth}, journalists are safe, to a certain extent, to embrace subjectivity in their profession, to craft and personalize stories in ways the journalists believe best engages their target audience. While this ritual may sustain the overall productivity of the newsroom, we wonder whether it truly empowers the individual visual data journalist as the ritual creates an undercurrent of skepticism towards creative expertise, narrative design, and other subjective approaches to appeal to an audience. This ritual asks those who partake in it to exert subtle forms of disciplinary power~\cite{foucault2003essential}\new{---promoting surveillance and productivity of the individual body as well as imposing specific norms and standards on the individual identity.} The availability of personal data today makes it ``ripe for exploitation''~\cite{cohen2019between}, and even well-meaning efforts to visualize data for granular insights can \new{risk privacy, exploitation, and discrimination}, particularly of socially vulnerable populations.

We see the mechanistic perspectives that are present in newsrooms \new{similarly: as strategic rituals that help journalists justify their design choices. At the same time, a purely mechanistic view of design creates cognitive dissonance when grappling with the emotional and ethical weight of data. We see participants in editorially flexible positions subvert this ritual.} Since much of visualization design language, e.g., statistics-driven design principles, pre-attentive features, mental models, is entrenched in Western mechanistic perspectives of human productivity, this departure from tradition is also a de-naturalization of Western views and an embrace of cultural and personal identity---and an opportunity to tell different kinds of data stories. 

Also entrenched in history is an implicit belief held among participants that mechanistic ways of knowing are constructed and learned. We wonder, to what extent, the humanistic ways of knowing are innate in contrast. Sociocultural factors influence visual perception, emotions, and metaphoric understanding~\cite{vcenvek2015cross, pokojna2024language}, and we see our participants use colour, symbols, and other aesthetics to leverage associations to, e.g., a group of scientists or a culture. We see an opportunity to examine how traditionally mechanistic languages of design can be re-imagined for humanistic insight through careful consideration of historical and sociocultural factors. 

\subsection{\new{Takeaways from thinking with critical theory}}

In contrast to traditional methods of inquiry, thinking explicitly with deconstruction enables a departure from structuralist systems of thought in visualization. \new{Reading for contradiction} subverts the endless ``repetition of what is known''~\cite{mazzei2021postqualitative}, normalizing a more critical \new{and expansive} interrogation of data that we collect (explicit) and do not collect (implicit). By examining the ideological reasons behind the emphasis or omission of certain perspectives, we \new{expose the hierarchal structure of beliefs that influence our ``truths'' about success and meaning in visual data journalism.} Yet, deconstructive reading ends in theory. Genealogical analysis brings theory into practice by contextualizing ideology as a product of historical events, material objects, and physical people and places. By tracing the evolution of ideas over time, we see the historical sweep of objectivity and mechanism that continues to exert power over data visualization practices today---constraining the truths data can or cannot tell, and the insights design can or cannot bring. 

\new{Recognizing the binaries in \textit{data as truth} can allow practitioners to deliberately adopt a different orientation about data as \textit{subjective} rather than \textit{objective} truth, e.g., model of data in a void ~\cite{ross2024almost}. Viewing data as plastic, for example, can inspire the development of visualization idioms, methods, and tools that intentionally create an environment for making sense of the data we have, e.g., interactive tables as a visualization idiom in their own right~\cite{bartram2021untidy}. Recognizing the binaries in \textit{design for insight} asks visualization practitioners to be cognizant of the virtues~\cite{correll2019ethical} we follow when we make decisions about the design and development of a visualization---in what we rationalize and normalize when repeating the mantra of \textit{visualization for insight}. For instance, the belief that visualizations can be ``too beautiful to be science'' (stated by D02)  articulates a longstanding bias associating aesthetics and appeal with deception, and reflects the cognitive dissonance practitioners may experience when drawing from humanistic, rather than purely mechanistic, values in developing visualizations for scientific insight.

By removing the ``blinders'' of objectivity and mechanism implicit in visualization, we allow ourselves to constructively criticize our own \textit{epistemology}. }We think about actionable research questions that can subvert the longstanding assumptions of \textit{data as truth} as well as the ways in which \textit{design (de)stabilizes such assumptions}. Consider the subversion of empiricism, for example: with a limited amount or complete absence of data, is it still ``scientifically productive'' to produce data-less visualizations, e.g., \textit{The Library of Missing Datasets}~\cite{onuoha2018missing}? What kinds of (new) domain knowledge might that produce? How would visualization epistemology---research questions, methods, frameworks, theories---need to change to allow for a new understanding of \new{``successful'' data visualization}? Finally, is it appropriate for visualization researchers to be responsible for such studies, or does the situation beg for collaboration with scholars trained in philosophy and literature?

Thinking with theory diversifies research outcomes by explicitly acknowledging and embracing the researchers' positionality. Part of the power of deconstruction is the flexibility it affords in interpretation---its reading is not attached to a particular ideology. While we focused on Western design philosophy, a new team of authors may trace a different genealogy based on a different set of binary oppositions. Researchers can further diverse research outcomes by pairing deconstruction with other methods, e.g., a Marxist reading might investigate the effect of ideology on economics and class politics in data journalism.

\section{Study Limitations}
\label{sec:limitations}

Although our interview corpus contains a rich trove of information, we only have space to unpack two sets of binary oppositions. Participants' detailed accounts may also not be generalizable to the broader visual data journalism community. While we engaged geographically diverse perspectives, the majority of participants are located in North American and European newsrooms. This in part informed our decision to cover Western design philosophy in our genealogy. We also acknowledge that about half of our participants were recruited from our personal network, as this potentially skews the perspectives in our work. The majority of our participants are not responsible for data stewardship e.g., collecting and preparing data for visualization. Recruiting participants with such responsibilities may have afforded us a deeper understanding of \textit{data as objective truth} by learning about the ways data are procured and processed in the newsroom.

\new{Given the flexibility of interpretation in deconstruction}, we wonder the extent to which our research outcomes would differ if we followed ideologies and histories privileged in non-Western perspectives. \new{Finally, a complete genealogy is not possible due to space limitations. We instead report a small set of relevant objects, ideas, or people that contextualize our deconstructive analysis findings. }

\section{Conclusion}
\label{sec:conclusion}

Visual data journalism as a medium concentrates the tensions broadly felt across visualization in the Information Age. Practitioners in this domain contend with the pervasive and centuries-long project of \textit{data as truth} that rests on foundational virtues such as objectivity and rationalism juxtaposed against subjectivity and the nature of craft. Following the visualization pipeline we localize \textit{design for insight}, a project resting on contradictory, historically-winding ideals that oppose the virtues of machine against human agency. We interviewed data journalists to understand the standards and beliefs that propel their sense of success and engagement in the work they produce. By examining their practice through a deconstructive lens we found a rich subtext of understanding in the things left unsaid. Rather than fixed in rigid structures of meaning, the concept of data malleability affords considered reflection on the rhetorical, crafted nature of data equal to its appearance of truth in the newsroom. Rethinking design as a vehicle for insight similarly asks one to reexamine assumptions of shared understanding and ways of seeing so as to consider new design patterns that blend mechanistic and humanistic ideals. 
Our work joins the voices of others in critical visualization research to encourage rethinking conceptions of data and design through new perspectives. By considering that these, and other notions considered central to engaging and successful journalism, are rather ``an effect of interpretation''~\cite{eco2000kant}, we may better identify our contingencies to make visible the complex and dynamic structures of the world in which we live. 

\section*{Supplemental Materials}
All supplemental materials are available on OSF (\href{https://creativecommons.org/licenses/by/4.0/}{CC BY 4.0}) at \href{https://osf.io/5fr48/}{\texttt{osf.io/5fr48}}. The repository includes our interview questions, group analysis board, and unedited interview quotes used in this paper.

\section*{Figure Credits}
\label{sec:figure_credits}

\cref{fig:teaser} is a collage created by the authors, using images credited under~\cref{fig:genealogy} imposed over Charles Darwin's Tree of primates, reproduced by kind permission of Cambridge Univ. Library (MS DAR 80; p.B91r). 
\cref{fig:genealogy} is a collage created by the authors: \textbf{A.} Leonardo da Vinci's Vitruvian Man, under public domain;
\textbf{B.} Athanasius Kircher's demonstration of the “great art of knowing”, under public domain;
\textbf{C.} Gerardus Mercator's 1569 world map, under public domain.
\textbf{D.} Bills of mortality, under public domain.
\textbf{E.} William Playfair’s trade-balance chart, under public domain.
\textbf{F.} Adriano Balbi and André-Michel Guerry’s Statistique comparée, under public domain.
\textbf{G.} Florence Nightingale’s Diagram of the causes of mortality in the army in the East, under public domain.
\textbf{H.} Charles Minard’s Carte figurative, under public domain.
\textbf{I.} Statistical graph, created by authors.
\textbf{J.} Nigel Holmes' Medical care expenditures, reproduced by kind permission of Nigel Holmes. 
\textbf{K.} The Gapminder World Map 2015, licensed under CC BY 4.0. 
\textbf{L.} Scrollytelling visualization, created by authors.
\textbf{M.} D3 visualizations © 2011 IEEE. Reprinted with permission from M. Bostock, V. Ogievetsky and J. Heer, ``D³ Data-Driven Documents,'' in \textit{IEEE Trans Vis Comput Graph}, 17(12):2301-2309, 2011. 
\textbf{N.} Epidemic curve, created by authors. 
\new{We as authors state \cref{fig:analysis,fig:objective-subjective,fig:humanism-mechanism} and~\cref{fig:genealogy}I, L and N are and remain under our own personal copyright, with permission to be used here. We also make them available under the \href{https://creativecommons.org/licenses/by/4.0/}{Creative Commons
Attribution 4.0 International (CC BY 4.0)} license and share them at \href{https://osf.io/5fr48/}{\texttt{osf.io/5fr48}}.}

\acknowledgments{%
	We are grateful to our participants for their time and insights that made this project possible. We are grateful to Wolfgang Hottner, Jill Walker Rettberg, Miriah Meyer, and Michael Correll for discussions and feedback on early drafts of this work. This work is supported by Univ. of Bergen and Trond Mohn Foundation (\#813558).%
}

\bibliographystyle{abbrv-doi-hyperref}

\bibliography{template}



\end{document}